\documentclass[AMA,STIX1COL]{WileyNJD-v2}
\usepackage{moreverb}
\usepackage{makecell}
\usepackage{multirow}
\usepackage{afterpage}

\newcommand\BibTeX{{\rmfamily B\kern-.05em \textsc{i\kern-.025em b}\kern-.08em
T\kern-.1667em\lower.7ex\hbox{E}\kern-.125emX}}

\articletype{research article}%

\received{<day> <Month>, <year>}
\revised{<day> <Month>, <year>}
\accepted{<day> <Month>, <year>}

\begin{document}

\title{Biomarker combination based on the Youden index with and without gold standard}

\author[1]{Ao Sun}

\author[2]{Yanting Li}

\author[3]{Xiao-Hua Zhou*}

\authormark{Sun \textsc{et al}}

\address[1]{\orgdiv{Center of Data Science}, \orgname{Peking University}, \orgaddress{\state{Beijing}, \country{China}}}

\address[2]{\orgdiv{Rheumatology Department}, \orgname{Guang'anmen Hospital}, \orgaddress{\state{Beijing}, \country{China}}}

\address[3]{\orgdiv{Department of Biostatistics and Beijing International Center for Mathematical Research}, \orgname{Peking University}, \orgaddress{\state{Beijing}, \country{China}}}

\corres{Xiao-Hua Zhou, Department of Biostatistics and Beijing International Center for Mathematical Research, Peking University, Beijing, China. \email{azhou@math.pku.edu.cn}}

\abstract[Abstract]{In clinical practice, multiple biomarkers are often measured on the same subject for disease diagnosis, and combining them can improve diagnostic accuracy. Existing studies typically combine multiple biomarkers by maximizing the Area Under the ROC Curve (AUC), assuming a gold standard exists or that biomarkers follow a multivariate normal distribution. However, practical diagnostic settings require both optimal combination coefficients and an effective cutoff value, and the reference test may be imperfect. In this paper, we propose a two-stage method for identifying the optimal linear combination and cutoff value based on the Youden index. First, it maximizes an approximation of the empirical AUC to estimate the optimal linear coefficients for combining multiple biomarkers. Then, it maximizes the empirical Youden index to determine the optimal cutoff point for disease classification. Under the semiparametric single index model and regularity conditions, the estimators for the linear coefficients, cutoff point, and Youden index are consistent. This method is also applicable when the reference standard is imperfect. We demonstrate the performance of our method through simulations and apply it to construct a diagnostic scale for Chinese medicine.}


\maketitle

\section{Introduction}\label{s:intro}

In fields like meteorology, economics, and computer science, leveraging multifaceted information for classification and prediction is crucial and has long been studied. Similarly, in medical practice, disease classification and prediction using clinical and laboratory data are key areas of research. Diagnostic studies often involve multiple tests on individuals, and no single test or biomarker is sufficient for precise diagnosis or prognosis. Therefore, integrating various sources of information is essential to enhance diagnostic accuracy \citep{schindler2021combining}.

Various methods exist for combining multiple biomarkers, with the simplest and most popular being linear combination. Pepe et al.\cite{pepe2006combining} employed likelihood maximization to estimate linear combination coefficients. Another common objective for linear combination is maximizing the area under the ROC curve (AUC), the most widely used index for summarizing the ROC curve, which evaluates the diagnostic accuracy of continuous or ordinal biomarkers \citep{zhou2011statistical}. Su and Liu\cite{su1993linear}, Reiser and Faraggi \cite{reiser1997confidence}, Liu et al. \citep{liu2005linear} optimized the AUC under the assumption that biomarkers follow a mixture of multivariate normal distributions. Pepe et al. \citep{pepe2006combining} relaxed the assumption of a parametric distribution, advocating for using the empirical AUC as the objective function and performing a grid search for the optimal linear combinations in a p-dimensional space. However, grid search becomes computationally challenging with more than three biomarkers, leading to the development of more efficient combination methods such as the min-max approach \citep{liu2011min}, stepwise method \citep{kang2013linear}, and approximation methods \citep{ma2007combining,huang2011optimal}.

While the AUC offers a comprehensive overview of diagnostic accuracy across all possible thresholds, a specific cutoff point is essential for diagnostic purposes. Surprisingly, very few research articles have explored combining biomarkers using alternative summary statistics of the ROC curve as objective functions beyond AUC. Some studies have numerically searched for optimal linear combinations that maximize sensitivity at a fixed specificity \citep{gao2008estimating,meisner2021combining}, while several methods have identified the best linear combination by maximizing partial AUC \citep{ma2019use,zhang2019feature,yan2018combining}.

The Youden index is another widely used summary index of the ROC curve\citep{youden1950index}. It represents the maximum of the sum of sensitivity and specificity minus one, ranging from 0 to 1, where 1 indicates complete separation of biomarker distributions in healthy and diseased populations, and 0 indicates complete overlap. In medical and biological sciences, it is extensively used for selecting diagnostic thresholds due to its seamless integration with the ROC framework \citep{perkins2006inconsistency} and its direct measurement of the maximum overall correct classification rate achievable by a biomarker. Combining biomarkers using the Youden index yields the highest possible overall correct classification rate at the diagnostic threshold among all feasible linear combinations, resulting in optimal diagnostic accuracy.

Existing methods for maximizing the Youden index often overlook the properties of the estimator and the inference process, focusing primarily on scenarios with a gold standard\citep{yin2014optimal,aznar2022stepwise,aznar2023comparing}. In practice, labels for diseased and healthy individuals are often based on imperfect reference tests, leading to potential misclassifications. Ignoring these misclassifications during diagnostic test development can severely bias accuracy parameter estimates \citep{lu2010bayesian}. To address the absence of a gold standard, latent-class models with two latent classes have been proposed. These models linearly combine continuous biomarkers to maximize AUC under the assumption of a multivariate normal distribution of biomarker values \citep{yu2011combining,garcia2016development,garcia2017estimation}. However, in real-world datasets, the assumption of a multivariate normal distribution is often too restrictive to be appropriate.

In this article, we introduce a two-stage method for estimating optimal combination coefficients and threshold, along with the corresponding Youden index. We establish the uniform consistency of the proposed estimators and propose a procedure to construct the confidence interval for the Youden index. Additionally, we demonstrate that our method achieves optimal combination coefficients and thresholds even when the reference standard is imperfect. To our knowledge, this is the first paper to address linear combination of biomarkers without parametric assumptions in the presence of an imperfect reference standard.

The remainder of the paper is organized as follows.  Sections \ref{s:notation} and \ref{existingmethods} introduce notations and provide a brief review of existing methods. Section \ref{proposed} presents the newly proposed two-stage combination method. Section \ref{imperfect} discusses the linear combination approach when the reference test is imperfect. Numerical studies, aimed at evaluating the performance of the method and applying it to a real data example, are presented in Section \ref{numerical}. Finally, Section \ref{conclusion} offers a discussion.

\section{Preliminaries}
\subsection{Notations}
\label{s:notation}
Consider a scenario involving $n$ patients, categorized into two groups: diseased and healthy. Let $D$ denote the presence ($D=1$) or absence ($D=0$) of the disease. There are $n_1$ diseased subjects and $n_0$ healthy subjects. Each patient is measured across $p$ biomarkers, denoted as $T_1, \cdots, T_p$. For the diseased group, let $T^1$ represent the p-dimensional vector of biomarker measurements, where $T_i^1=(T_{i1}^1,T_{i2}^1,\cdots, T_{ip}^1)$ denotes the observed biomarker values for the $i$th diseased subject $(i=1,\cdots,n_1)$. Here, $T_{ik}^1$ is the value of the $k$th biomarker for the $i$th individual. Similarly, for the healthy group, $T_j^0=(T_{j1}^0,T_{j2}^0,\cdots, T_{jp}^0)$ represents the biomarker values for the $j$th healthy subject $(j=1,\cdots,n_0)$, and $T_{jk}^0$ is the measurement of the $k$th biomarker for the $j$th healthy individual.

We consider linear combination scores of the form $\omega_0^{\prime}T=T_1+\cdots+\omega_{0p} T_p$, where $\omega_0=(1,\omega_{02},\cdots,\omega_{0p})$ is the vector of combination coefficients. This linear score omits an intercept, and the coefficient for $T_1$ is set to 1 without loss of generality. If $\omega_{01}>0$, a linear predictor of the form $\gamma+\omega_{01}\cdot\omega_0^{\prime}T$ exceeding a threshold is equivalent to using $\omega_0^{\prime}T$. Additionally, we can redefine $T_1$ as $-T_1$ to ensure $\omega_{01}>0$, covering all possible cases. Since the receiver operating characteristic (ROC) curves for $\omega_0^{\prime}T$ and $\gamma+\omega_{01}\cdot\omega_0^{\prime}T$ are identical, it suffices to focus on the linear combination $\omega_0^{\prime}T$. If the relationship between $T$ and $D$ can be described using a semiparametric single index model, the probability of disease can be expressed as: $\mathbb{P}(D=1|T)=H(\omega_0^{\prime}T)$, where $H$ is an unknown increasing link function. According to Neyman and Pearson \citep{neyman1933ix}, the combination score $\omega_0^{\prime}T$ yields the optimal ROC curve, meaning no other combination score based on $T$ can outperform it at any accuracy point $(\text{FPR},\text{TPR})$ on the ROC curve. Therefore, for a fixed false positive rate (FPR), the true positive rate (TPR) for the rule $\omega_0^{\prime}T>c$ is higher than that of any other score with the same FPR. The Youden index, which quantifies the maximum vertical distance between the ROC curve and the diagonal $y=x$ \citep{schisterman2005optimal}, is defined as:
\begin{equation}
    J_{\omega_0}=\max_c\left\{Se_{\omega_0}(c)+Sp_{\omega_0}(c)-1\right\}=\max_c\left\{F_{0\omega_0}(c)-F_{1\omega_0}(c)\right\},
\end{equation}
where $\text{Se}_{\omega_0}(c)=1-F_{1\omega_0}(c)$ is the sensitivity, and $\text{Sp}_{\omega_0}(c)=F_{0\omega_0}(c)$ is the specificity at threshold $c$. Here, $F_{1\omega_0}(\cdot)$ and $F_{0\omega_0}(\cdot)$ denote the cumulative distribution functions of $\omega_0^{\prime}T^1$ and $\omega_0^{\prime}T^0$, respectively, while $f_{0\omega_0}(\cdot)$ and $f_{1\omega_0}(\cdot)$ are their corresponding densities. Given the optimality of the ROC curve, the score $\omega_0^{\prime}T$ achieves the maximum Youden index among all possible combination scores.

We typically assume that the supports of $\omega_0^{\prime}T^1$ and $\omega_0^{\prime}T^0$ overlap. Moreover, we assume the existence of a threshold $c_0$ such that $f_{1\omega_0}(c_0)=f_{0\omega_0}(c_0)$, with $f_{0\omega_0}(t)<f_{1\omega_0}(t)$ for $t<c_0$ and $f_{0\omega_0}(t)>f_{1\omega_0}(t)$  for $t>c_0$. This condition holds, for instance, when the likelihood ratio is monotonic. It implies that $\omega_0^{\prime}T^0$ is stochastically smaller than $\omega_0^{\prime}T^1$, i.e. $F_{1\omega_0}(t)\geq F_{0\omega_0}(t)$ for all $t$. Under this assumption, the sum $Se_{\omega_0}(c)+Sp_{\omega_0}(c)$ is maximized by selecting $c=c_0$. The Youden index for the linear combination score $\omega_0^{\prime}T$, denoted as $J_{\omega_0}$, then can be given by:
\begin{equation}
    J_{\omega_0}=F_{0\omega_0}(c_0)-F_{1\omega_0}(c_0).
\end{equation}
Our primary objective is to estimate the optimal combination coefficients $\omega_0$ that maximize the Youden index, the corresponding cutoff point $c_0$, and the maximum Youden index $J_{\omega_0}$.

\subsection{Existing methods}
\label{existingmethods}
Existing methods for finding optimal linear combination of multiple biomarkers to maximize the Youden index can be categorized into empirical searching methods and derivation-based numerical searching methods. These methods lack well-defined estimator properties and do not address scenarios where the reference test is imperfect. 

\subsubsection{Empirical searching methods}
Given $\omega$, denote the optimal cutoff point as $c_{\omega}$. The empirical estimate of the Youden index $J_{\omega}$ is:
\begin{equation}\label{obj}
    \widehat{J}_{\omega}=\frac{\sum_{j=1}^{n_0}\mathbb{I}\left(\omega^{\prime}T_j^0\leq c_{\omega}\right)}{n_0}-\frac{\sum_{i=1}^{n_1}\mathbb{I}\left(\omega^{\prime}T_i^1\leq c_{\omega}\right)}{n_1}.
\end{equation}
Optimizing this non-continuous objective function via grid search for multiple biomarkers has a computational cost of $\mathcal{O}(n^p)$, where $n$ is the sample size and $p$ is the number of biomarkers. 

Two empirical searching methods, the min-max method and the stepwise method, have been proposed to address this computational challenge \citep{yu2011combining}. The min-max method simplifies the optimization by only considering the minimum and maximum values of the biomarker measurements, reducing the $p$-dimensional grid search to a single coefficient $\omega$. However, a key limitation of the min-max approach is its reliance solely on the minimum and maximum values of the biomarkers, which may not result in the optimal linear combination. 

The stepwise approach reduces computational complexity by initially calculating the empirical Youden index for each biomarker and ranking them. It starts by combining the two biomarkers with the highest Youden indices, using grid search to optimize the coefficients, then sequentially adds additional biomarkers. While effective, this stepwise method is not guaranteed to find the global optimum and can be cumbersome when dealing with binary biomarkers.

\subsubsection{Derivation-based numerical searching methods}
There are two derivation-based methods for numerically searching for the optimal linear combination maximizing the Youden index \citep{yu2011combining}. One method is parametric, assuming biomarkers follow a multivariate normal distribution in both diseased and healthy states. Under this assumption, explicit solutions for the optimal cutoff value and Youden index can be derived. However, this assumption may not hold in real-world scenarios.

The other method is non-parametric, using an approximation for the indicator function in objective function (\ref{obj}) and simultaneously maximizing it with respect to $\omega$ and $c$. The estimator $(\widehat{\omega}^{\text{SIM}},\widehat{c}^{\text{SIM}})$ is defined:
\begin{equation*}
    (\widehat{\omega}^{\text{SIM}},\widehat{c}^{\text{SIM}})=\arg\max_{\omega,c}\left\{\frac{1}{n_0}\sum_{j=1}^{n_0}\Phi\left(\frac{c-\omega^{\prime}T_j^0}{h}\right)-\frac{1}{n_1}\sum_{i=1}^{n_1}\Phi\left(\frac{c-\omega^{\prime}T_i^1}{h}\right)\right\}.
\end{equation*}
However, the coefficients $\omega$ and the cutoff value $c_{\omega}$ are highly correlated, making simultaneous optimization with quasi-Newton algorithm prone to converging to local minima\citep{wright2006numerical}.

\section{The proposed method}
\label{proposed}
In this section, we propose a two-stage method to estimate the optimal linear coefficients and cutoff point, establish the consistency of the corresponding estimators, and introduce a new confidence interval for the optimal Youden index.

\subsection{Estimation}
Given that under the single index model, the optimal combination score with the largest Youden index also has the largest AUC, we propose a two-stage approach to estimate the coefficients $\omega_0$ and the cutoff point $c_0$. In the first stage, we aim to maximize the empirical AUC to estimate the optimal linear combination coefficients $\omega_0$. The empirical AUC of a linear risk score, $\omega^{\prime}T$, is defined as:
\begin{equation}
    \widehat{\text{AUC}}(\omega)=\frac{\sum\limits_{i=1}^{n_1}\sum\limits_{j=1}^{n_0}\left\{\mathbb{I}\left[\omega^{\prime}T_i^1>\omega^{\prime}T_j^0\right]+\frac{1}{2}\mathbb{I}\left[\omega^{\prime}T_i^1=\omega^{\prime}T_j^0\right]\right\}}{n_1n_0},
\end{equation}
where $\mathbb{I}(\cdot)$ is the indicator function. This objective function is discontinuous due to the indicator function, necessitating brute force search or specialized algorithms. The computational cost grows exponentially with $n^p$, making optimization based on $\widehat{\text{AUC}}(\omega)$ impractical for large numbers of biomarkers.

To address this, we approximate the indicator function with a differentiable function. According to Lin et al.\citep{lin2011selection}, using the standard normal distribution function for this approximation is more accurate and stable than using the sigmoid function. Therefore, we use $\Phi\left(\left(\omega^{\prime}T_i^1-\omega^{\prime}T_j^0\right)/h_n\right)$ as a smooth approximation to $\mathbb{I}\left[\omega^{\prime}T_i^1>\omega^{\prime}T_j^0\right]+\frac{1}{2}\mathbb{I}\left[\omega^{\prime}T_i^1=\omega^{\prime}T_j^0\right]$, where $\Phi(\cdot)$ is the cumulative distribution function of the standard normal variable and $h_n$ is a bandwidth chosen to converge to zero, specifically $h_n=(n_1n_0)^{-0.1}$, as recommended by Vexler et al. \citep{vexler2006note} The resulting estimator, denoted as $\widehat{\omega}$, is defined as:
\begin{equation}\label{estimateomega}
    \widehat{\omega}=\arg\max_{\omega}\left\{\frac{1}{n_1n_0}\sum\limits_{i=1}^{n_1}\sum\limits_{j=1}^{n_0}\Phi\left(\frac{\omega^{\prime}T_i^1-\omega^{\prime}T_i^0}{h}\right)\right\}.
\end{equation}
The estimator is identifiable up to a scale constant if at least one component of $T$ is continuous.

After obtaining the estimator of coefficients, the optimal cutoff point is estimated by:
\begin{equation}\label{hac}
\widehat{c}^{\text{TS}}=\text{median}\left\{c:\max_c\frac{\sum_{j=1}^{n_0}\mathbb{I}\left(\widehat{\omega}^{\prime}T_j^0\leq c\right)}{n_0}-\frac{\sum_{i=1}^{n_1}\mathbb{I}\left(\widehat{\omega}^{\prime}T_i^1\leq c\right)}{n_1}\right\}.
\end{equation}
Alternatively, the maximum or minimum value instead of the median can be used in (\ref{hac}).

The estimated Youden index, denoted as $\widehat{J}^{\text{TS}}$, is given by:
\begin{equation}
    \widehat{J}^{\text{TS}}=\frac{\sum_{j=1}^{n_0}\mathbb{I}\left(\widehat{\omega}^{\prime}T_j^0\leq \widehat{c}^{\text{TS}}\right)}{n_0}-\frac{\sum_{i=1}^{n_1}\mathbb{I}\left(\widehat{\omega}^{\prime}T_i^1\leq \widehat{c}^{\text{TS}}\right)}{n_1}.
\end{equation}

\subsection{Asymptotic properties}
The asymptotic properties of $\widehat{\omega}$ have been proven by Ma and Huang \citep{ma2007combining}. For completeness, we briefly review these properties. Let $\theta_0=(\omega_{02},\cdots,\omega_{0p})$, and $\widehat{\theta}_n=(\widehat{\omega}_2,\cdots,\widehat{\omega}_p)$. For the consistency of $\widehat{\theta}_n$, we assume the following conditions: (A1) The true parameter value $\theta_0$ is an interior point of $\theta$, which is a compact subset of $\mathbb{R}^{p-1}$. (A2) Let $S_T$ denote the support of the biomarker vector $T$, where $S_T$ is not contained in any proper linear subspace of $\mathbb{R}^p$ and the first component of $T$ has an everywhere positive density, conditional on the other components. 

\begin{lemma}\label{lemma}
    Suppose assumptions $(A1)$ and $(A2)$ hold and $h_n\rightarrow 0$ as $n\rightarrow\infty$. Then $\widehat{\omega}\rightarrow_{P}\omega_0$ as $n\rightarrow\infty$.
\end{lemma}

For the consistency of $\widehat{c}^{\text{TS}}$ and $\widehat{J}^{\text{TS}}$, we assume: (B) For any $\delta>0$, there exists $\epsilon>0$, such that $\sup_{|x-c_0|>\delta}\left[F_{0\omega_0}(x)-F_{1\omega_0}(x)\right] < F_{0\omega_0}(c_0)-F_{1\omega_0}(c_0)-\epsilon$. This condition is slightly weaker than the assumption in Section \ref{s:notation} and ensures that the maximum of $F_{0\omega_0}(x)-F_{1\omega_0}(x)$ occurs at $c_0$ and is separated from other values by at least $\epsilon$.

\begin{theorem}\label{theorem1}
    Suppose assumptions (A1), (A2) and (B) hold and $h_n \rightarrow 0$ as $n \rightarrow \infty$. Then $\widehat{c}^{\text{TS}}\rightarrow_{P}c_0$ and $\widehat{J}^{\text{TS}}\rightarrow_{P}J_{\omega_0}$ as $n\rightarrow\infty$.
\end{theorem}
Theorem 1 provides asymptotic consistency for estimators of the cutoff point and Youden index, with the proof presented in Appendix \ref{app1}.

\subsection{Inference for Youden index}
According to Shan \citep{shan2015improved}, the coverage property of Wald-type confidence intervals for the Youden index is generally unsatisfactory and we use the square-and-add limits based on the Wilson score method to construct the non-parametric confidence interval for the Youden index $J_{\omega_0}$.

For simplicity, denote $p_1=P(T^1\leq c_0)$ and $p_0=P(T^0\leq c_0)$. The Wilson confidence intervals for $p_1$, denoted as $(l_1, u_1)$, are the roots of the following equality:
\begin{equation*}
    (p_1-\widehat{p}_1)^2=z_{1-\alpha/2}^2\frac{p_1(1-p_1)}{n_1}.
\end{equation*}
From this, we can calculate:
\begin{equation*}
    l_1=\frac{1}{1+z_{1-\alpha/2}^2/n_1}\left[\widehat{p}_1+\frac{z_{1-\alpha/2}^2}{2n_1}-z_{1-\alpha/2}\sqrt{\frac{\widehat{p}_1(1-\widehat{p}_1)}{n_1}+\frac{z_{1-\alpha/2}^2}{4n_1^2}}\right],
\end{equation*}
and
\begin{equation*}
    u_1=\frac{1}{1+z_{1-\alpha/2}^2/n_1}\left[\widehat{p}_1+\frac{z_{1-\alpha/2}^2}{2n_1}+z_{1-\alpha/2}\sqrt{\frac{\widehat{p}_1(1-\widehat{p}_1)}{n_1}+\frac{z_{1-\alpha/2}^2}{4n_1^2}}\right].
\end{equation*}
The confidence intervals for $\text{Var}(\widehat{p}_1)$ can be estimated by $\widehat{\text{Var}}_l(\widehat{p}_1)=l_1(1-l_1)/n_1$, $\widehat{\text{Var}}_u(\widehat{p}_1)=u_1(1-u_1)/n_1$.

The Wilson confidence interval for $p_0$ can be calculated similarly and denoted as $(l_0,u_0)$. The confidence intervals for $\text{Var}(\widehat{p}_0)$ can be estimated as $\widehat{\text{Var}}_l(\widehat{p}_0)=l_0(1-l_0)/n_0$ and $\widehat{\text{Var}}_u(\widehat{p}_0)=u_0(1-u_0)/n_0$.

Assuming independence of test results between the healthy and diseased groups, after obtaining $\widehat{\omega}$ and $\widehat{c}^{\text{TS}}$, the variance of $\widehat{J}^{\text{TS}}$ can be consistently estimated as $\widehat{\text{Var}}(\widehat{J}^{\text{TS}})=\widehat{\text{Var}}(\widehat{p}_0-\widehat{p}_1)=\widehat{\text{Var}}(\widehat{p}_0)+\widehat{\text{Var}}(\widehat{p}_1)$.

We use the AC estimators for $p_1$ and $p_0$, as proposed by Agresti and Choll \citep{agresti1998approximate}, denoted as $\widehat{p}_1$ and $\widehat{p}_0$, which are as follows:
\begin{equation*}
    \widehat{p}_0=\frac{\sum_{j=1}^{n_0}\mathbb{I}\left(\widehat{\omega}^{\prime}T_j^0\leq \widehat{c}\right)+\frac{1}{2}z_{1-\alpha/2}^2}{n_0+z_{1-\alpha/2}^2},\; \widehat{p}_1=\frac{\sum_{i=1}^{n_1}\mathbb{I}\left(\widehat{\omega}^{\prime}T_i^1\leq \widehat{c}\right)+\frac{1}{2}z_{1-\alpha/2}^2}{n_1+z_{1-\alpha/2}^2}.
\end{equation*}
And the adjusted estimator for the Youden index is defined as:
\begin{equation*}
    \widehat{J}^{\text{AC}}=\frac{\sum_{j=1}^{n_0}\mathbb{I}\left(\widehat{\omega}^{\prime}T_j^0\leq \widehat{c}\right)+\frac{1}{2}z_{1-\alpha/2}^2}{n_0+z_{1-\alpha/2}^2}-\frac{\sum_{i=1}^{n_1}\mathbb{I}\left(\widehat{\omega}^{\prime}T_i^1\leq \widehat{c}\right)+\frac{1}{2}z_{1-\alpha/2}^2}{n_1+z_{1-\alpha/2}^2},
\end{equation*}
According to Zhou and Qin\citep{zhou2012new}, the adjusted estimator $\widehat{J}^{\text{AC}}$ has smaller bias than the empirical estimator $\widehat{J}^{\text{TS}}$ when $J\leq 0.8$ and when sample size are small. Therefore, we propose calculating the confidence interval of $J_0$ as:
\begin{equation*}
    (J_L, J_U)=\left(\widehat{J}^{\text{AC}}-z_{1-\alpha/2}\sqrt{\widehat{\text{Var}}_l(\widehat{p}_0)+\widehat{\text{Var}}_u(\widehat{p}_1)},\widehat{J}^{\text{AC}}+z_{1-\alpha/2}\sqrt{\widehat{\text{Var}}_u(\widehat{p}_0)+\widehat{\text{Var}}_l(\widehat{p}_1)}\right).
\end{equation*}
This confidence interval does not require bootstrap sampling, making it computationally efficient.

\section{When the reference test is imperfect}
\label{imperfect}
In numerous scenarios, while the true disease status, or gold standard, have a well-defined clinical definition, practical constraints can render it unobservable or latent. Instead, we observe an imperfect reference test that is prone to errors \citep{dawid1979maximum,walter1988estimation}. Existing methods for combining biomarkers, when only imperfect gold standards are available, assume a multivariate normal distribution for the joint distribution of biomarkers and focus on maximizing the AUC rather than the Youden index \citep{yu2011combining,garcia2016development,garcia2017estimation}. We will identify the optimal linear combination and cutoff point that maximizes the Youden index without making parametric assumptions. 

Denote the imperfect reference test as $R$. The positive predictive value (ppv) of $R$ is defined as $\text{ppv}=\mathbb{P}(D=1|R=1)$ and the negative predictive value (npv) is defined as $\text{npv}=\mathbb{P}(D=0|R=0)$. We make two general assumptions:
\begin{assumption}\label{assum1}
    $ppv+npv>1.$
\end{assumption}
\begin{assumption}\label{assum2}
    Given the true disease status, biomarkers and the imperfect reference test are independent.
\end{assumption}
Assumption \ref{assum1} indicates that the reference test is more accurate than random guessing, while Assumption \ref{assum2} suggests that biomarkers and the reference standard are improbable to misdiagnose the same patients.

When considering the imperfect test $R$ as the reference, for a linear combination $\omega^{\prime}T$, we denote the corresponding proxy AUC as $\widetilde{AUC}_{\omega}$, which is defined as:
\begin{equation}
    \begin{split}
        \widetilde{\text{AUC}}(\omega)&=P(\omega^{\prime}T_i>\omega^{\prime}T_j|R_i=1,R_j=0),
    \end{split}
\end{equation}
and its relationship with $\text{AUC}(\omega)$, which is the true $\text{AUC}$ for score $\omega^{\prime}T$, can be expressed as:
\begin{equation}\label{relationship1}
    \widetilde{\text{AUC}}(\omega)=(ppv+npv-1)\text{AUC}(\omega)-\frac{1}{2}(ppv+npv)+1.
\end{equation}
Similarly, we denote the Youden index when $R$ is regarded as the reference as $\widetilde{J}_{\omega}$, defined as:
\begin{equation}
    \begin{split}
        \widetilde{J}_{\omega}&=\max_c\left\{P(\omega^{\prime}T>c|R=1)+P(\omega^{\prime}T<c|R=0)-1\right\}, 
    \end{split}
\end{equation}
and its relationship with $J_{\omega}$, which is the true Youden index, can be expressed as:
\begin{equation}\label{relationship2}
    \widetilde{J}_{\omega}=(ppv+npv-1)J_{\omega}.
\end{equation}

Under assumption \ref{assum1}, $\widetilde{\text{AUC}}(\omega)$ and $\widetilde{J}_{\omega}$ increase monotonically with $\text{AUC}(\omega)$ and $J_{\omega}$, respectively. Consequently, the combination coefficients $\omega_0$ that maximize $\text{AUC}(\omega)$ also maximize $\widetilde{\text{AUC}}(\omega)$. Similarly, the cutoff point $c_0$ that achieves $J_{\omega_0}$ also achieves the maximum $\widetilde{J}_{\omega_0}$. The derivation of (\ref{relationship1}) and (\ref{relationship2}) can be found in Appendix \ref{app2}.

\subsection{Estimation}
Assume there are $\Tilde{n}_1$ subjects with diseased imperfect reference result and $\Tilde{n}_0$ subjects with healthy imperfect reference result. Denote the biomarker measurements, whose result of imperfect reference standard is diseased, as $\widetilde{T}^1$, and $\widetilde{T}^0$ as the measurements with healthy reference result. Denote the proxy AUC estimator as $\Breve{\omega}$, which is calculated as:
\begin{equation}\label{breveomega}
    \Breve{\omega}=\arg\max_{\omega}\left\{\frac{1}{\Tilde{n}_1\Tilde{n}_0}\sum\limits_{i=1}^{\Tilde{n}_1}\sum\limits_{j=1}^{\Tilde{n}_0}\Phi\left(\frac{\omega^{\prime}\widetilde{T}_i^1-\omega^{\prime}\widetilde{T}_i^0}{h}\right)\right\}.
\end{equation}
This estimator is also only identifiable up to a scale constant if at least one component of $T$ is continuous.

After obtaining the estimator of coefficients, the cutoff point is estimated by
\begin{equation}\label{brevec}
\Breve{c}=\text{median}\left\{c:\max_c\frac{\sum_{j=1}^{\Tilde{n}_0}\mathbb{I}\left(\Breve{\omega}^{\prime}\widetilde{T}_j^0\leq c\right)}{\Tilde{n}_0}-\frac{\sum_{i=1}^{\Tilde{n}_1}\mathbb{I}\left(\Breve{\omega}^{\prime}\widetilde{T}_i^1\leq c\right)}{\Tilde{n}_1}\right\}.
\end{equation}
Alternatively, we can use the maximum or minimum value instead of the median in (\ref{brevec}).

The estimated proxy Youden index, denoted as $\Breve{J}$, is given by
\begin{equation}
    \Breve{J}=\frac{\sum_{j=1}^{\Tilde{n}_0}\mathbb{I}\left(\Breve{\omega}^{\prime}\widetilde{T}_j^0\leq \Breve{c}\right)}{\Tilde{n}_0}-\frac{\sum_{i=1}^{\Tilde{n}_1}\mathbb{I}\left(\Breve{\omega}^{\prime}\widetilde{T}_i^1\leq \Breve{c}\right)}{\Tilde{n}_1}.
\end{equation}

\subsection{Asymtotic properties}
For identifiability, we also assume the first component of $\omega_0$ is equal to 1. The asymptotic properties of $\Breve{\omega}$ can be seen a simple extension to Lemma \ref{lemma}.

\begin{theorem}
    Suppose assumptions $(A1)$ and $(A2)$ hold and $h_n\rightarrow 0$ as $n\rightarrow\infty$. Then $\Breve{\omega}\rightarrow_{P}\omega_0$ as $n\rightarrow\infty$.
\end{theorem}

The consistency of $\Breve{c}$ and $\Breve{J}$ can also be seen as a simple extension to Theorem \ref{theorem1}. We assume that: (C) For any $\delta>0$, there exists $\epsilon>0$, such that $\sup_{|x-c_0|>\delta}\left[\widetilde{F}_{0\omega_0}(x)-\widetilde{F}_{1\omega_0}(x)\right] < \widetilde{F}_{0\omega_0}(c_0)-\widetilde{F}_{1\omega_0}(c_0)-\epsilon$, where $\widetilde{F}_{0\omega_0}(x)$ is defined as $\mathbb{P}(\omega^{\prime}T\leq x|R=0)$ and $\widetilde{F}_{1\omega_0}(x)$ is defined as $\mathbb{P}(\omega^{\prime}T\leq x|R=1)$.

\begin{theorem}
    Suppose assumption (A1), (A2) and (C) hold, and $h_n \rightarrow 0$ as $n \rightarrow \infty$. Then $\Breve{c}\rightarrow_{P}c_0$ and $(ppv+npv-1)\Breve{J}\rightarrow_{P}J_{\omega_0}$ as $n\rightarrow\infty$.
\end{theorem}

In conclusion, when only an imperfect reference test is available and the two general assumptions hold, we can effectively treat the imperfect reference test as the gold standard. We can then utilize the proposed two-stage method to determine the optimal coefficients and cutoff value.

\section{Numerical studies}
\label{numerical}
This section comprises four parts: Section \ref{simu1} presents the coverage probabilities and interval lengths of the proposed confidence interval. Section \ref{simu2} compares our proposed two-stage method with the simultaneous optimization approach under both perfect and imperfect reference tests. Section \ref{simu3} evaluates the performance of the two-stage method relative to the simultaneous optimization method across scenarios where the single-index model holds and where it does not. Lastly, Section \ref{simu4} demonstrates the application of our proposed method to real-world data.

\subsection{Coverage probabilities and lengths of confidence intervals}\label{simu1}
We assess the coverage probabilities and average lengths of the proposed confidence interval and demonstrate its effectiveness by generating 1000 random samples with a total size of $n$. Specially, we drew samples of size $\pi\times n$ from the distribution function for test results of diseased patients and another independent sample of size $(1-\pi)\times n$ from the distribution function for healthy subjects. The sample size $n$ were set to 100, 200 and 400, with prevalence $\pi$ set to 0.25, 0.5 and 0.75, respectively. Through simulation, we compare the coverage rate of our confidence interval with that of the NP confidence interval proposed by Shan \citep{shan2015improved}, which uses the same variance estimators but empirical Youden index estimator.

In our study, multivariate normal distributions were chosen as the underlying distributions. Specifically, we selected $T^0\sim MVN \left(0,I_5\right)$ for healthy subjects and $T^1\sim MVN \left(\mu_1,I_5\right)$ for diseased subjects. The optimal linear coefficient is $\omega_0=\mu_1$, and $\mu_1$ was chosen to achieve predefined Youden index values of 0.45, 0.60, 0.70, and 0.85. The results are summarized in Table \ref{confidenceinterval}. 

From the results, it is evident that as the sample size increases, the length of the confidence interval decreases. In balanced data scenarios ($\pi=0.50$), the confidence interval is narrower compared to unbalanced data scenarios ($\pi=0.25$ or $0.75$). Additionally, the larger the true Youden index value, the narrower the confidence interval. Compared to the previously used NP confidence interval, our proposed confidence interval generally has higher coverage rates, consistently achieving the nominal 95\% level.

\begin{table*}[]
    \centering
    \caption{Coverage rates (CR), average lengths (AL), lower limit (LL), upper limit (UL) of our newly proposed 95\% confidence intervals for the Youden index, as well as the coverage rates (CR\_NP) of the 95\% confidence intervals using the NP method based on 1000 replications.}\label{confidenceinterval}
    {\begin{tabular*}{\textwidth}{@{\extracolsep\fill}llllllll@{\extracolsep\fill}}
        \toprule
        \textbf{Youden index value} & $\mathbf{n_1}$ &$\mathbf{n_0}$ & \textbf{CR\_NP} & \textbf{CR} & \textbf{AL} & \textbf{LL} & \textbf{UL}  \\
        \midrule
         $J_0= 0.45$ & 50 & 50 & 0.854 & 0.944 & 0.3225 & 0.3257 & 0.6482  \\

         &100 & 100 & 0.890 & 0.948 & 0.2352 & 0.3586 & 0.5938   \\
         
         &200 & 200 & 0.902 & 0.930 & 0.1692 & 0.3837 & 0.5530  \\

         &75 & 25 & 0.851 & 0.963 & 0.3643 & 0.3032 & 0.6674 \\
         
         &150 & 50 & 0.885 & 0.937 & 0.2662 & 0.3446 & 0.6108 \\

         &300 & 100 & 0.898 & 0.936 & 0.1933 & 0.3772 & 0.5705 \\
         
         &25 & 75 & 0.848 & 0.965 & 0.3616 & 0.3054 & 0.6671 \\

         &50 & 150 & 0.890 & 0.961 & 0.2649 & 0.3431 & 0.6080 \\
         
         &100 & 300 & 0.903 & 0.951 & 0.1922 & 0.3763 & 0.5685 \\

         \multicolumn{6}{@{}l}{ } \\

         $J_0=0.60$ & 50 & 50 & 0.888 & 0.989 & 0.2941 & 0.4583 & 0.7524 \\

         &100 & 100 & 0.896 & 0.972 & 0.2119 & 0.5052 & 0.7171 \\
         
         &200 & 200 & 0.918 & 0.955 & 0.1521 & 0.5350 & 0.6871 \\

         &75 & 25 & 0.886 & 0.981 & 0.3359 & 0.4318 & 0.7677 \\
         
         &150 & 50 & 0.905 & 0.977 & 0.2425 & 0.4870 & 0.7295 \\

         &300 & 100 & 0.902 & 0.955 & 0.1734 & 0.5273 & 0.7007 \\
         
         &25 & 75 & 0.894 & 0.984 & 0.3326 & 0.4339 & 0.7665 \\

         &50 & 150 & 0.885 & 0.968 & 0.2404 & 0.4871 & 0.7275 \\
         
         &100 & 300 & 0.912 & 0.950 & 0.1730 & 0.5250 & 0.6979 \\

         \multicolumn{6}{@{}l}{ } \\
           
         $J_0=0.70$ & 50 & 50 & 0.929 & 0.978 & 0.2694 & 0.5486 & 0.8179 \\

         &100 & 100 & 0.906 & 0.970 & 0.1920 & 0.6002 & 0.7921 \\
         
         &200 & 200 & 0.914 & 0.962 & 0.1362 & 0.6369 & 0.7731 \\

         &75 & 25 & 0.927 & 0.992 & 0.3116 & 0.5181 & 0.8297 \\
         
         &150 & 50 & 0.920 & 0.970 & 0.2214 & 0.5806 & 0.8020 \\

         &300 & 100 & 0.924 & 0.967 & 0.1568 & 0.6240 & 0.7808 \\
         
         &25 & 75 & 0.937 & 0.984 & 0.3096 & 0.5153 & 0.8249 \\

         &50 & 150 & 0.914 & 0.972 & 0.2182 & 0.5840 & 0.8021 \\
         
         &100 & 300 & 0.922 & 0.969 & 0.1564 & 0.6230 & 0.7793 \\

         \multicolumn{6}{@{}l}{ } \\

         $J_0=0.85$ & 50 & 50 & 0.979 & 0.986 & 0.2191 & 0.6905 & 0.9096 \\

         &100 & 100 & 0.952 & 0.978 & 0.1492 & 0.7531 & 0.9022 \\
         
         &200 & 200 & 0.955 & 0.976 & 0.1037 & 0.7885 & 0.8922 \\

         &75 & 25 & 1.000 & 0.963 & 0.2664 & 0.6453 & 0.9117 \\
         
         &150 & 50 & 0.964 & 0.969 & 0.1778 & 0.7270 & 0.9048 \\

         &300 & 100 & 0.947 & 0.972 & 0.1208 & 0.7761 & 0.8969 \\
         
         &25 & 75 & 0.999 & 0.963 & 0.2645 & 0.6464 & 0.9109 \\

         &50 & 150 & 0.963 & 0.968 & 0.1757 & 0.7290 & 0.9048 \\
         
         &100 & 300 & 0.954 & 0.962 & 0.1201 & 0.7745 & 0.8947 \\ 
         \bottomrule
    \end{tabular*}}
    \bigskip
\end{table*}

\subsection{Investigation of performance with perfect and imperfect reference tests}
\label{simu2}
We conducted simulation studies to evaluate the performance of our newly proposed two-stage method under perfect and imperfect reference tests when the single index model is satisfied. The Youden index was predefined as 0.45 and 0.70, with sample sizes of 200, 400, and 800 and prevalences of 0.25, 0.5 and 0.75. These were randomly split into training and testing sets with equal sample size. The sensitivity and specificity of the reference test were set to 1, 0.95, 0.90, 0.85, respectively. For each scenario, 1000 random samples were generated.

The results are displayed in Table \ref{imperfectreference} and indicate that as the sensitivity and specificity of the reference test decrease, performance deteriorates in both training and testing sets. As the sample size increases, the Youden index values approach the true value $J_0$ in both the training and testing set. And in all scenarios, our method outperforms the simultaneous optimization method, with the advantage becoming more pronounced as the reference test's accuracy decreases. 

\begin{table*}[]
    \caption{The performance of our newly-proposed two-stage method (TSM) and existing simultaneous optimization method (SIM) when the reference standard is perfect and imperfect}\label{imperfectreference}
    {\begin{tabular*}{\textwidth}{@{\extracolsep\fill}llllllllllll@{\extracolsep\fill}}
        \toprule
        & & & & \multicolumn{2}{@{}l}{\textbf{Se=Sp=1}} & \multicolumn{2}{@{}l}{\textbf{Se=Sp=0.95}} & \multicolumn{2}{@{}l}{\textbf{Se=Sp=0.90}} & \multicolumn{2}{@{}l}{\textbf{Se=Sp=0.85}} \\\cmidrule{5-6}\cmidrule{7-8}\cmidrule{9-10}\cmidrule{11-12}
        
        $\mathbf{J_0}$ & $\mathbf{\pi}$ & $\mathbf{n}$ & \textbf{Method} & \textbf{Train} & \textbf{Test} & \textbf{Train} & \textbf{Test} & \textbf{Train} & \textbf{Test} & \textbf{Train} & \textbf{Test} \\
        \midrule
         $0.45$ & $0.50$ & 200 & TSM & 0.5527 & 0.3901 & 0.5280 & 0.3753 & 0.5043 & 0.3602 & 0.4692 & 0.3340  \\

         & & & SIM & 0.4928 & 0.2872 & 0.4656 & 0.2744 & 0.4337 & 0.2639 & 0.4033 & 0.2440 \\
         
         & & 400 & TSM & 0.5125 & 0.4172 & 0.5010 & 0.4104 & 0.4881 & 0.4034 & 0.4696 & 0.3884 \\

         & & & SIM & 0.4683 & 0.3233 & 0.4419 & 0.3096 & 0.4226 & 0.2981 & 0.3955 & 0.2841 \\
         
         & & 800 & TSM & 0.4842 & 0.4270 & 0.4776 & 0.4249 & 0.4701 & 0.4197 & 0.4608 & 0.4132 \\

         & & & SIM & 0.4416 & 0.3491 & 0.4310 & 0.3457 & 0.4144 & 0.3325 & 0.3959 & 0.3243 \\
         
         & $0.75$ & 200 & TSM & 0.5695 & 0.3817 & 0.5374 & 0.3612 & 0.4997 & 0.3403 & 0.4564 & 0.3142  \\

         & & & SIM & 0.5087 & 0.2702 & 0.4734 & 0.2554 & 0.4390 & 0.2487 & 0.3993 & 0.2242 \\
         
         & & 400 & TSM & 0.5274 & 0.4084 & 0.5093 & 0.4007 & 0.4855 & 0.3817 & 0.4571 & 0.3616 \\

         & & & SIM & 0.4782 & 0.3047 & 0.4530 & 0.2953 & 0.4237 & 0.2813 & 0.3899 & 0.2644 \\
         
         & & 800 & TSM & 0.4951 & 0.4234 & 0.4852 & 0.4166 & 0.4696 & 0.4092 & 0.4553 & 0.3988 \\

         & & & SIM & 0.4520 & 0.3367 & 0.4356 & 0.3280 & 0.4135 & 0.3165 & 0.3906 & 0.3024 \\

         & $0.25$ & 200 & TSM & 0.5811 & 0.3754 & 0.5420 & 0.3539 & 0.5019 & 0.3307 & 0.4556 & 0.3047 \\

         & & & SIM & 0.5148 & 0.2702 & 0.4720 & 0.2513 & 0.4319 & 0.2336 & 0.3850 & 0.2115 \\
         
         & & 400 & TSM & 0.5235 & 0.4084 & 0.5056 & 0.3984 & 0.4856 & 0.3868 & 0.4620 & 0.3716 \\

         & & & SIM & 0.4720 & 0.3144 & 0.4439 & 0.2993 & 0.4125 & 0.2848 & 0.3814 & 0.2663 \\
         
         & & 800 & TSM & 0.4905 & 0.4239 & 0.4805 & 0.4198 & 0.4675 & 0.4094 & 0.4551 & 0.3996 \\

         & & & SIM & 0.4487 & 0.3434 & 0.4273 & 0.3318 & 0.4060 & 0.3191 & 0.3784 & 0.2978 \\

         \multicolumn{11}{@{}c}{} \\
         
         $0.70$ & $0.50$ & 200 & TSM & 0.7761 & 0.6603 & 0.7543 & 0.7292 & 0.6468 & 0.6295 & 0.6986 & 0.6038  \\

         & & & SIM & 0.6917 & 0.5239 & 0.6593 & 0.5000 & 0.6137 & 0.4744 & 0.5724 & 0.4400 \\
         
         & & 400 & TSM & 0.7430 & 0.6792 & 0.7330 & 0.6721 & 0.7179 & 0.6618 & 0.7028 & 0.6513 \\

         & & & SIM & 0.6922 & 0.5843 & 0.6657 & 0.5665 & 0.6398 & 0.5453 & 0.6075 & 0.5228 \\
         
         & & 800 & TSM & 0.7259 & 0.6905 & 0.7204 & 0.6861 & 0.7128 & 0.6818 & 0.7044 & 0.6757 \\

         & & & SIM & 0.7046 & 0.6415 & 0.6890 & 0.6293 & 0.6681 & 0.6137 & 0.6470 & 0.5944 \\
         
         & $0.75$ & 200 & TSM & 0.7855 & 0.6491 & 0.7515 & 0.6294 & 0.7180 & 0.6131 & 0.6721 & 0.5683  \\

         & & & SIM & 0.6909 & 0.4997 & 0.6537 & 0.4651 & 0.6167 & 0.4495 & 0.5712 & 0.4184 \\
         
         & & 400 & TSM & 0.7541 & 0.6684 & 0.7343 & 0.6557 & 0.7074 & 0.6393 & 0.6793 & 0.6187 \\

         & & & SIM & 0.6899 & 0.5476 & 0.6600 & 0.5311 & 0.6239 & 0.5011 & 0.5823 & 0.4711 \\
         
         & & 800 & TSM & 0.7335 & 0.6855 & 0.7227 & 0.6772 & 0.7091 & 0.6687 & 0.6912 & 0.6539  \\

         & & & SIM & 0.7017 & 0.6155 & 0.6784 & 0.5991 & 0.6518 & 0.5773 & 0.6172 & 0.5482 \\

         & $0.25$ & 200 & TSM & 0.7906 & 0.6409 & 0.7430 & 0.6201 & 0.7127 & 0.5906 & 0.6602 & 0.5530  \\

         & & & SIM & 0.6975 & 0.4974 & 0.6370 & 0.4617 & 0.5852 & 0.4264 & 0.5363 & 0.3921 \\
         
         & & 400 & TSM & 0.7558 & 0.6740 & 0.7370 & 0.6609 & 0.7152 & 0.6485 & 0.6912 & 0.6300 \\

         & & & SIM & 0.7041 & 0.5830 & 0.6636 & 0.5523 & 0.6163 & 0.5194 & 0.5783 & 0.4867 \\
         
         & & 800 & TSM & 0.7334 & 0.6844 & 0.7239 & 0.6786 & 0.7121 & 0.6680 & 0.6954 & 0.6562 \\

         & & & SIM & 0.7062 & 0.6266 & 0.6753 & 0.6014 & 0.6440 & 0.5750 & 0.6080 & 0.5471 \\ 
         \bottomrule
    \end{tabular*}}
    \bigskip
\end{table*}

\subsection{Comparison of the proposed method and existing method}
\label{simu3}
We also conducted simulation studies to compare the performance of our newly proposed two-stage method with the existing simultaneous optimization method under scenarios where the single index model is satisfied and not satisfied. We present results for two scenarios: multivariate normal distribution with equal covariance matrix and unequal covariance matrices. When the covariance matrices differ between diseased and healthy populations, the single index model is not satisfied. 

In the scenario with equal covariance matrices, diseased and healthy samples were drawn from multivariate normal distributions with mean vectors $\mu_0=(0,0,0,0,0)^{\prime}$ and $\mu_1=(0.4,0.7,1.0,1.3,1.6)^{\prime}$. Both groups shared the same covariance matrix, $\Sigma_1=\Sigma_0=(1-\gamma)I_5+\gamma J_5$, where $I_5$ is the identity matrix, $J_5$ is a matrix of all ones, and $\gamma$ was set to 0.3, 0.5, and 0.7. In the scenario with unequal covariance matrices, diseased and healthy samples were drawn from multivariate normal distributions with the same mean vectors but different covariance matrices: $\Sigma_1=0.3I_5+0.7J_5$ and $\Sigma_0=0.7I_5+0.3J_5$. 

For both settings, prevalences were set to 0.25, 0.5 and 0.75. We generated 1000 random samples for each simulation setting, with sample size of $n=200$, $n=400$, and $n=800$. Each dataset was randomly split into equal-sized training and testing sets. The coefficients, mean and variance of the estimated Youden index in the training and testing sets are presented in Tables \ref{equalcovariance} and \ref{unequalcovariance}.

The simulation results demonstrate that our newly proposed method performs well in both scenarios, whether the single index model holds (Table \ref{equalcovariance}) or not (Table \ref{unequalcovariance}). Our two-stage method consistently achieves higher Youden index values compared to the simultaneous optimization method. Furthermore, our method exhibits lower variance, indicating greater stability in performance.

\begin{table*}[]
    \caption{The performance of our newly-proposed two-stage method (TSM) and existing simultaneous optimization method (SIM) with samples from multivariate normal distributions with equal covariance matrices}\label{equalcovariance}
    {\begin{tabular*}{\textwidth}{@{\extracolsep\fill}llllllllll@{\extracolsep\fill}}
        \toprule
         & & \multicolumn{4}{@{}l}{\textbf{Estimated Youden index}} & \multicolumn{4}{@{}l}{\textbf{Variance of Youden index}}  \\\cmidrule{3-6}\cmidrule{7-10}

         & & \multicolumn{2}{@{}l}{\textbf{Training set}} & \multicolumn{2}{@{}l}{\textbf{Testing set}} & \multicolumn{2}{@{}l}{\textbf{Training set}} & \multicolumn{2}{@{}l}{\textbf{Testing set}} \\\cmidrule{3-4}\cmidrule{5-6}\cmidrule{7-8}\cmidrule{9-10}

         \textbf{Prevalence} & \textbf{Sample size} & \textbf{TSM} & \textbf{SIM} & \textbf{TSM} & \textbf{SIM} & \textbf{TSM} & \textbf{SIM} & \textbf{TSM} & \textbf{SIM} \\
        \midrule
         \multicolumn{10}{@{}c}{$\mathbf{\Sigma}_1=\mathbf{\Sigma}_0=0.3*\boldsymbol{I}_5+0.7*\boldsymbol{J}_5$ (large correlation)} \\
         
         $\pi=0.5$ & $n=200$ & 0.7696 & 0.7443 & 0.6570 & 0.6141 & 0.0041 & 0.0048 & 0.0064 & 0.0081 \\

         &$n=400$ & 0.7402 & 0.7303 & 0.6744 & 0.6465 & 0.0019 & 0.0024 & 0.0029 & 0.0039 \\
         
         &$n=800$ & 0.7249 & 0.7194 & 0.6862 & 0.6623 & 0.0011 & 0.0014 & 0.0014 & 0.0019\\

         $\pi=0.75$ & $n=200$ & 0.7893 & 0.7561 & 0.6506 & 0.5944 & 0.0052 & 0.0072 & 0.0093 & 0.0124 \\

         &$n=400$ & 0.7513 & 0.7370 & 0.6718 & 0.6364 & 0.0025 & 0.0031 & 0.0044 & 0.0062 \\
         
         &$n=800$ & 0.7291 & 0.7225 & 0.6819 & 0.6518 & 0.0014 & 0.0017 & 0.0019 & 0.0026\\

         $\pi=0.25$ & $n=200$ & 0.7850 & 0.7617 & 0.6467 & 0.6152 & 0.0045 & 0.0052 & 0.0100 & 0.0117 \\

         &$n=400$ & 0.7562 & 0.7442 & 0.6683 & 0.6373 & 0.0027 & 0.0030 & 0.0041 & 0.0050 \\
         
         &$n=800$ & 0.7330 & 0.7255 & 0.6826 & 0.6552 & 0.0015 & 0.0017 & 0.0019 & 0.0025\\

         \multicolumn{10}{@{}c}{$\mathbf{\Sigma}_1=\mathbf{\Sigma}_0=0.5*\boldsymbol{I}_5+0.5*\boldsymbol{J}_5$ (medium correlation)} \\
         
         $\pi=0.5$ & $n=200$ & 0.7279 & 0.7003 & 0.6009 & 0.5485 & 0.0043 & 0.0053 & 0.0072 & 0.0096 \\

         &$n=400$ & 0.6939 & 0.6795 & 0.6214 & 0.5819 & 0.0024 & 0.0032 & 0.0035 & 0.0052 \\
         
         &$n=800$ & 0.6743 & 0.6692 & 0.6353 & 0.6049 & 0.0013 & 0.0019 & 0.0017 & 0.0026\\

         $\pi=0.75$ & $n=200$ & 0.7485 & 0.7070 & 0.5951 & 0.5283 & 0.0060 & 0.0070 & 0.0097 & 0.0140 \\

         &$n=400$ & 0.7034 & 0.6860 & 0.6160 & 0.5666 & 0.0030 & 0.0040 & 0.0046 & 0.0074 \\
         
         &$n=800$ & 0.6813 & 0.6726 & 0.6297 & 0.5916 & 0.0016 & 0.0025 & 0.0024 & 0.0036\\

         $\pi=0.25$ & $n=200$ & 0.7434 & 0.7155 & 0.5918 & 0.5432 & 0.0052 & 0.0058 & 0.0106 & 0.0127 \\

         &$n=400$ & 0.7055 & 0.6900 & 0.6163 & 0.5713 & 0.0029 & 0.0040 & 0.0047 & 0.0062 \\
         
         &$n=800$ & 0.6846 & 0.6785 & 0.6289 & 0.5990 & 0.0017 & 0.0024 & 0.0021 & 0.0033\\

         \multicolumn{10}{@{}c}{$\mathbf{\Sigma}_1=\mathbf{\Sigma}_0=0.7*\boldsymbol{I}_5+0.3*\boldsymbol{J}_5$ (low correlation)} \\
         
         $\pi=0.5$ & $n=200$ & 0.7342 & 0.6946 & 0.6097 & 0.5340 & 0.0043 & 0.0060 & 0.0071 & 0.0109 \\

         &$n=400$ & 0.7008 & 0.6785 & 0.6332 & 0.5755 & 0.0022 & 0.0041 & 0.0035 & 0.0073 \\
         
         &$n=800$ & 0.6812 & 0.6714 & 0.6402 & 0.6052 & 0.0012 & 0.0023 & 0.0017 & 0.0036\\

         $\pi=0.75$ & $n=200$ & 0.7544 & 0.7045 & 0.5972 & 0.5050 & 0.0054 & 0.0073 & 0.0090 & 0.0145 \\

         &$n=400$ & 0.7094 & 0.6795 & 0.6226 & 0.5560 & 0.0027 & 0.0046 & 0.0044 & 0.0081 \\
         
         &$n=800$ & 0.6855 & 0.6698 & 0.6368 & 0.5896 & 0.0016 & 0.0033 & 0.0022 & 0.0047\\

         $\pi=0.25$ & $n=200$ & 0.7425 & 0.7015 & 0.6043 & 0.5295 & 0.0049 & 0.0071 & 0.0106 & 0.0142 \\

         &$n=400$ & 0.7136 & 0.6914 & 0.6250 & 0.5689 & 0.0029 & 0.0047 & 0.0040 & 0.0079 \\
         
         &$n=800$ & 0.6899 & 0.6778 & 0.6375 & 0.5980 & 0.0017 & 0.0031 & 0.0020 & 0.0043\\  

         \bottomrule
    \end{tabular*}}
    \bigskip
\end{table*}

\begin{table*}[]
    \caption{The performance of our newly-proposed two-stage method (TSM) and existing simultaneous optimization method (SIM) with samples from multivariate normal distributions with unequal covariance matrices}\label{unequalcovariance}
    {\begin{tabular*}{\textwidth}{@{\extracolsep\fill}llllllllll@{}}
        \toprule
         & & \multicolumn{4}{@{}l}{\textbf{Estimated Youden index}} & \multicolumn{4}{@{}l}{\textbf{Variance of Youden index}}  \\\cmidrule{3-6}\cmidrule{7-10}

         & & \multicolumn{2}{@{}l}{\textbf{Training set}} & \multicolumn{2}{@{}l}{\textbf{Testing set}} & \multicolumn{2}{@{}l}{\textbf{Training set}} & \multicolumn{2}{@{}l}{\textbf{Testing set}} \\\cmidrule{3-4}\cmidrule{5-6}\cmidrule{7-8}\cmidrule{9-10}

         \textbf{Prevalence} & \textbf{Sample size} & \textbf{TSM} & \textbf{SIM} & \textbf{TSM} & \textbf{SIM} & \textbf{TSM} & \textbf{SIM} & \textbf{TSM} & \textbf{SIM} \\
        \midrule
         $\pi=0.5$ & $n=200$ & 0.7302 & 0.7042 & 0.6043 & 0.5584 & 0.0045 & 0.0052 & 0.0069 & 0.0089 \\

         &$n=400$ & 0.6968 & 0.6842 & 0.6191 & 0.5855 & 0.0022 & 0.0027 & 0.0034 & 0.0047 \\
         
         &$n=800$ & 0.6744 & 0.6694 & 0.6341 & 0.6048 & 0.0016 & 0.0011 & 0.0016 & 0.0024\\

        \multicolumn{10}{@{}c}{ } \\
         
         $\pi=0.75$ & $n=200$ & 0.7507 & 0.7192 & 0.5971 & 0.5322 & 0.0056 & 0.0058 & 0.0093 & 0.0129 \\

         &$n=400$ & 0.7064 & 0.6934 & 0.6160 & 0.5725 & 0.0029 & 0.0034 & 0.0050 & 0.0073 \\
         
         &$n=800$ & 0.6828 & 0.6761 & 0.6290 & 0.5944 & 0.0017 & 0.0019 & 0.0021 & 0.0033\\

         \multicolumn{10}{@{}c}{ } \\

         $\pi=0.25$ & $n=200$ & 0.7390 & 0.7139 & 0.5998 & 0.5532 & 0.0049 & 0.0052 & 0.0101 & 0.0112 \\

         &$n=400$ & 0.7056 & 0.6937 & 0.6210 & 0.5820 & 0.0029 & 0.0035 & 0.0043 & 0.0052 \\
         
         &$n=800$ & 0.6815 & 0.6758 & 0.6320 & 0.6030 & 0.0015 & 0.0019 & 0.0022 & 0.0028\\  

         \bottomrule
    \end{tabular*}}
    \bigskip
\end{table*}

We conducted additional simulations with all biomarkers being binary, where the cutoff value and linear coefficients may not be unique, to assess whether our method can achieve a higher Youden index and thereby evaluate its robustness. Five biomarkers were generated following a Bernoulli distribution with parameters $p=(0.7,0.6,0.5,0.4,0.3)$ respectively. The risk score was modeled using logistic regression, where coefficients were chose to achieve prevalences of $0.25$, $0.5$ and $0.75$. For each scenario, sample sizes of 200, 400 and 800 were used, with samples equally divided into training and testing sets. The results are presented in Table \ref{webtable1}. The findings indicate that even with all biomarkers being binary, our method demonstrates superior performance with lower variance compared to the simultaneous optimization method. As the sample size increases, the Youden index in the testing sets improves while variance decreases.
\begin{table*}[]
    \caption{The performance of our newly-proposed two-stage method (TSM) and existing simultaneous optimization method (SIM) with all biomarkers being binary}\label{webtable1}
    {\begin{tabular*}{\columnwidth}{@{\extracolsep\fill}llllllllll@{}}
        \toprule
         & & \multicolumn{4}{@{}l}{\textbf{Estimated Youden index}} & \multicolumn{4}{@{}l}{\textbf{Variance of Youden index}}  \\\cmidrule{3-6}\cmidrule{7-10}

         & & \multicolumn{2}{@{}l}{\textbf{Training set}} & \multicolumn{2}{@{}l}{\textbf{Testing set}} & \multicolumn{2}{@{}l}{\textbf{Training set}} & \multicolumn{2}{@{}l}{\textbf{Testing set}} \\\cmidrule{3-4}\cmidrule{5-6}\cmidrule{7-8}\cmidrule{9-10}

         \textbf{Prevalence} & \textbf{Sample size} & \textbf{TSM} & \textbf{SIM} & \textbf{TSM} & \textbf{SIM} & \textbf{TSM} & \textbf{SIM} & \textbf{TSM} & \textbf{SIM} \\
        \midrule

        \multicolumn{10}{@{}c}{$\omega=(-2.25,-0.80,0.65,2.10,3.55)$} \\
        
         $\pi=0.5$ & $n=200$ & 0.6475 & 0.6202 & 0.5495 & 0.5241 & 0.0046 & 0.0052 & 0.0086 & 0.0101 \\

         &$n=400$ & 0.6235 & 0.5983 & 0.5688 & 0.5479 & 0.0026 & 0.0028 & 0.0041 & 0.0044 \\
         
         &$n=800$ & 0.6149 & 0.5921 & 0.5818 & 0.5622 & 0.0014 & 0.0016 & 0.0019 & 0.0022\\

        \multicolumn{10}{@{}c}{$\omega=(-1.00, 0.10, 1.20, 2.30, 3.40)$} \\
         
         $\pi=0.75$ & $n=200$ & 0.6031 & 0.5597 & 0.4913 & 0.4503 & 0.0072 & 0.0075 & 0.0126 & 0.0150 \\

         &$n=400$ & 0.5781 & 0.5390 & 0.5159 & 0.4806 & 0.0044 & 0.0046 & 0.0072 & 0.0071 \\
         
         &$n=800$ & 0.5675 & 0.5256 & 0.5318 & 0.4968 & 0.0028 & 0.0030 & 0.0038 & 0.0038\\

         \multicolumn{10}{@{}c}{$\omega=(-2.40, -1.30, -0.20, 0.90, 2.00)$} \\

         $\pi=0.25$ & $n=200$ & 0.6178 & 0.5853 & 0.4786 & 0.4501 & 0.0065 & 0.0070 & 0.0125 & 0.0142 \\

         &$n=400$ & 0.5828 & 0.5540 & 0.5011 & 0.4776 & 0.0036 & 0.0041 & 0.0058 & 0.0063 \\
         
         &$n=800$ & 0.5632 & 0.5371 & 0.5206 & 0.4978 & 0.0019 & 0.0022 & 0.0028 & 0.0032\\  
         \bottomrule
    \end{tabular*}}
    \bigskip
\end{table*}

\subsection{Data example}
\label{simu4}
Rheumatoid arthritis (RA) is a chronic autoimmune disease that primarily affects joints and surrounding tissues, causing inflammation, joint damage, disability, and work loss \cite{gravallese2023rheumatoid}. RA patients face increased mortality risks from cardiovascular diseases, respiratory conditions, and infections compared to the general population \cite{turk2023non}. Pharmacological treatments, including various forms of disease-modifying antirheumatic drugs (DMARDs) and glucocorticoids, aim to achieve sustained remission or low disease activity \cite{smolen2023eular}. However, some patients experience poor responses or adverse effects, necessitating exploration of alternative therapies \cite{nagy2021eular}.

Nowadays, traditional Chinese medicine (TCM) is increasingly recognized worldwide, with evidence suggesting that integrative TCM and Western medicine can improve treatment outcomes compared to Western medicine alone \cite{xing2020efficacy}. In TCM, syndrome differentiation (ZHENG) is central to treatment, functioning as a diagnostic framework guided by TCM theory. However, syndrome differentiation relies heavily on practitioners' subjective judgment, influenced by clinical experience and training. This variability often results in diagnostic inconsistencies, with reported inter-practitioner agreement as low as 30\% \cite{zhang2008improvement}. To address this, standardized diagnostic scales for various conditions, including coronary heart disease, heart failure, and ulcerative colitis, have been developed \cite{fang2019development, leung2023validation, chen2018development}.

RA has been a key focus of traditional Chinese medicine (TCM) for over 2000 years, rooted in classical texts like the Inner Canon of Huangdi (Huangdi Neijing). Damp-Heat Impeding Syndrome (DHIS) is a common TCM pattern in RA, affecting 43.86\% of cases \cite{wang2018multi}. Network meta-analyses suggest that herbs targeting DHIS can reduce inflammation, alleviate symptoms, and improve drug tolerance \cite{li2021comparative}. Diagnostic scales for DHIS have been developed, but they primarily rely on the Delphi method, which collects expert opinions through multiple rounds of anonymous surveys \cite{graham2003delphi}. This method depends solely on expert consensus without incorporating data-driven analysis, limiting its reliability. The latest DHIS diagnostic scale yields a Youden index of just 0.28, with a specificity of 1 but a sensitivity of only 0.28, indicating overly conservative performance.\cite{jiang2018guidelines}

In this study, we construct a diagnostic scale for Damp-Heat Impeding Syndrome (DHIS) using our newly proposed method. We utilize data from the China Rheumatoid Arthritis Registry of Patients with Chinese Medicine (CERTAIN), the first prospective, multicenter registry for RA patients receiving TCM treatment \cite{gong2022china}. CERTAIN includes over 14,000 RA patients from 145 centers across 30 provinces in China. Our analysis focuses on 4,310 patients with clinician-confirmed diagnoses and complete records of 27 symptoms relevant to DHIS, along with two supporting laboratory indicators. To ensure data accuracy, clinical symptoms—including tongue and pulse manifestations—were initially recorded by a junior physician and subsequently reviewed and corrected by a senior physician. Syndrome diagnoses were independently made by an associate chief physician or higher. Therefore, it is reasonable to assume that the recorded symptoms and syndrome diagnoses may introduce independent errors in classification \cite{wang2016evaluation}.

We split the dataset into a training set (70\%) and a testing set (30\%) to compare the performance of our proposed two-stage method with the existing simultaneous optimization method for constructing TCM scales to diagnose dampness-heat obstruction syndrome. The results, summarized in Table \ref{realdata}, include the optimal Youden index, sensitivity, and specificity for each method. The two-stage method achieves a Youden index of 0.69 on the training set—20\% higher than the 0.58 achieved by the simultaneous optimization method. Both methods yield similar Youden indices on the testing set, indicating the stability of the two-stage approach and the absence of overfitting.

\begin{table*}[]
    \caption{Data for diagnosing dampness-heat obstruction syndrome.}\label{realdata}
    {\begin{tabular*}{\columnwidth}{@{\extracolsep\fill}lllllll@{}}
        \toprule
        &\multicolumn{2}{@{}l}{\textbf{Estimated Youden index}} & \multicolumn{2}{@{}l}{\textbf{Sensitivity}} & \multicolumn{2}{@{}l}{\textbf{Specificity}} \\\cmidrule{2-3}\cmidrule{4-5}\cmidrule{6-7}
         \textbf{Method} & \textbf{Training} & \textbf{Testing} & \textbf{Training} & \textbf{Testing} & \textbf{Training} & \textbf{Testing} \\
         \midrule
         TSM & 0.6924 & 0.6928 & 0.7235 & 0.7158 & 0.9688 & 0.9769 \\
         SIM & 0.5783 & 0.5728 & 0.8333 & 0.8183 & 0.7450 & 0.7544 \\
        \bottomrule
    \end{tabular*}}
    \bigskip
\end{table*}

\section{discussion}
\label{conclusion}
In this article, we introduce a novel method for linearly combining multiple biomarkers to maximize the Youden index under both perfect and imperfect reference standard scenarios. Our approach operates in two stages: first optimizing an empirical AUC approximation to determine optimal linear coefficients, and then maximizing the empirical Youden index to find the optimal disease classification cutoff. Under the semiparametric single index model and certain regularity conditions, our method provides consistent estimators for the linear coefficients, cutoff point, and Youden index. Furthermore, we propose a confidence interval for the Youden index using the AC estimator and Wilson score method, demonstrating robust coverage rates and computational efficiency.

Additionally, we used simulations to compare the performance of our newly proposed two-stage method with the existing simultaneous optimization method. Simulations were conducted under scenarios where the single-index model holds and where it does not. Furthermore, we evaluated the robustness of our method by performing simulations with all biomarkers set as binary. Such binary biomarker settings are common in diagnostic and prognostic assays, particularly in traditional Chinese medicine. Through simulations and a real data example, we demonstrated that our proposed method outperforms the existing approach in nearly all scenarios considered. Notably, when all biomarkers are binary, our method also performs well, with the estimated Youden index exhibiting strong performance on both training and testing datasets. Extensive simulation studies further validate the performance of our method and assess its effectiveness across various conditions.

However, we observe that the performance of the Youden index in the testing set consistently lags behind that in the training set for both methods. Improving generalization remains an open research challenge. Additionally, the assumption of a single index model may not universally hold. In clinical settings, it may not be practical to combine all biomarkers available, especially when some provide little information. Including such biomarkers could even degrade the performance of the combination score's Youden index. Hence, our future research will focus on relaxing the single index model assumption and devising methods to select an optimal subset of biomarkers for efficient and effective linear combination in practical applications.

\newpage


\newpage

\appendix
\section{Proof of Theorem 1}\label{app1}
In our article, the true optimal cutoff value $c_0$ is defined as:
\begin{equation}
    c_0=\arg\max_{c}\left\{F_{0\omega_0}(c)-F_{1\omega_0}(c)\right\},
\end{equation}
and the optimal Youden index is defined as:
\begin{equation}
    J_{\omega_0}=\max_c\left\{Se_{\omega_0}(c)+Sp_{\omega_0}(c)-1\right\}=F_{0\omega_0}(c_0)-F_{1\omega_0}(c_0).
\end{equation} 

Define the empirical distribution functions as follows:
\begin{equation*}
\begin{split}
    &\widehat{F}_{0\widehat{\omega}}(c)=\frac{1}{n_0}\sum_{j=1}^{n_0}\mathbb{I}\left(\widehat{\omega}^{\prime}T_j^0\leq c\right), \\
    & \widehat{F}_{0\omega_0}(c)=\frac{1}{n_0}\sum_{j=1}^{n_0}\mathbb{I}\left(\omega_0^{\prime}T_j^0\leq c\right), \\
    & F_{0\omega_0}(c)=\mathbb{P}\left(\omega_0^{\prime}T^0\leq c\right).
\end{split}
\end{equation*}
Similarly, define $\widehat{F}_{1\widehat{\omega}}(c)$, $\widehat{F}_{1\omega_0}(c)$, and $F_{1\omega_0}(c)$ analogously. Our estimator is defined as:
\begin{equation}\label{hatc}
    \widehat{c}^{\text{TS}}=\text{median}\left\{c:\max_c\widehat{F}_{0\widehat{\omega}}(c)-\widehat{F}_{1\widehat{\omega}}(c)\right\},
\end{equation}
\begin{equation}
    \widehat{J}^{\text{TS}}=\widehat{F}_{0\widehat{\omega}}(\widehat{c}^{\text{TS}})-\widehat{F}_{1\widehat{\omega}}(\widehat{c}^{\text{TS}}).
\end{equation}
Alternatively, the maximum or minimum value instead of the median can be used in (\ref{hatc}).

First, we aim to prove that for any given $c$, $\widehat{F}_{0\widehat{\omega}}(c)\stackrel{P}{\rightarrow}F_{0\omega_0}(c)$ and $\widehat{F}_{1\widehat{\omega}}(c)\stackrel{P}{\rightarrow}F_{1\omega_0}(c)$ as $n\rightarrow\infty$.

Using the triangle inequality:
\begin{equation}\label{ineq1}
    \left\lvert\widehat{F}_{0\widehat{\omega}}(c)-F_{0\omega_0}(c)\right\rvert \leq \left\lvert\widehat{F}_{0\widehat{\omega}}(c)-\widehat{F}_{0\omega_0}(c)\right\rvert+\left\lvert\widehat{F}_{0\omega_0}(c)-F_{0\omega_0}(c)\right\rvert.
\end{equation}
According to Glivenko-Cantelli theorem, 
\begin{equation*}
    \sup_c\left\lvert\widehat{F}_{0\omega_0}(c)-F_{0\omega_0}(c)\right\rvert\stackrel{a.s.}{\rightarrow}0.
\end{equation*}
Next we will prove that the first term on the right-hand side of inequality (\ref{ineq1}) also converges to 0 in probability.

Since $\widehat{\omega}\stackrel{P}{\rightarrow}\omega_0$, for any $\epsilon>0$, for any $\delta>0$, there exists $N$, such that when $n>N$,
\begin{equation*}
    \mathbb{P}\left(\Vert\widehat{\omega}-\omega_0\Vert<\epsilon\right)>1-\delta.
\end{equation*}
Since $\omega$ is a finite-dimensional vector, we have
\begin{equation*}
    \vert\widehat{\omega}^{\prime}T-\omega_0^{\prime}T\vert\leq \Vert\widehat{\omega}-\omega_0\Vert \Vert T\Vert.
\end{equation*}
For any $\epsilon>0$, define:
\begin{equation*}
    A_{\epsilon}=\left\{T:\vert\widehat{\omega}^{\prime}T-\omega_0^{\prime}T\vert<\epsilon\right\},
\end{equation*}
and we have $\mathbb{P}(A_{\epsilon})\rightarrow1$. That is, for any given c, for any $\delta>0$, $\epsilon>0$, there exists $N$, such that when $n>N$, 
\begin{equation}
    \mathbb{P}\left(\left\lvert\mathbb{I}(\widehat{\omega}^{\prime}T\leq c)-\mathbb{I}(\omega_0^{\prime}T\leq c)\right\rvert>\epsilon\right)<\delta.
\end{equation}
Then we have $\left\lvert\widehat{F}_{0\widehat{\omega}}(c)-\widehat{F}_{0\omega_0}(c)\right\rvert\stackrel{P}{\rightarrow}0$ as $n\rightarrow \infty$.

Similarly, we can prove that $\left\lvert\widehat{F}_{1\widehat{\omega}}(c)-\widehat{F}_{1\omega_0}(c)\right\rvert\stackrel{P}{\rightarrow}0$ as $n\rightarrow\infty$.

Using condition (B), let $\epsilon^{\prime}=\frac{\epsilon}{5}$. From the convergence of $\widehat{F}_{0\widehat{\omega}}$ and $\widehat{F}_{1\widehat{\omega}}$, for large enough $n_0$ and $n_1$, we have
\begin{equation*}
    F_{0\omega_0}(c)-F_{1\omega_0}(c)-2\epsilon^{\prime} < \widehat{F}_{0\widehat{\omega}}(c)-\widehat{F}_{1\widehat{\omega}}(c) < F_{0\omega_0}(c)-F_{1\omega_0}(c)+2\epsilon^{\prime}\quad \text{for}\;\text{all}\; c.
\end{equation*}
Hence
\begin{equation*}
    \begin{split}
        \sup_{|c-c_0|>\delta}\left[\widehat{F}_{0\widehat{\omega}}(c)-\widehat{F}_{1\widehat{\omega}}(c)\right] & \leq 2\epsilon^{\prime}+\sup_{|c-c_0|>\delta}\left(F_{0\omega_0}(c)-F_{1\omega_0}(c)\right) \\
        &<F_{0\omega_0}(c_0)-F_{1\omega_0}(c_0)-3\epsilon^{\prime} \quad \text{by}\;\text{condition}\;(B) 
    \end{split}
\end{equation*}
Because
\begin{equation*}
    \widehat{F}_{0\widehat{\omega}}(c_0)-\widehat{F}_{1\widehat{\omega}}(c_0)\stackrel{P}{\rightarrow}F_{0\omega_0}(c_0)-F_{1\omega_0}(c_0),
\end{equation*}
for large enough $n_0$ and $n_1$, we have:
\begin{equation*}
    \widehat{F}_{0\widehat{\omega}}(c_0)-\widehat{F}_{1\widehat{\omega}}(c_0)>F_{0\omega_0}(c_0)-F_{1\omega_0}(c_0)-\epsilon^{\prime}.
\end{equation*}
Thus,
\begin{equation*}
    \sup_{|c-c_0|>\delta}\left[\widehat{F}_{0\widehat{\omega}}(c)-\widehat{F}_{1\widehat{\omega}}(c)\right]<\widehat{F}_{0\widehat{\omega}}(c_0)-\widehat{F}_{1\widehat{\omega}}(c_0).
\end{equation*}
Therefore,
\begin{equation*}
    \sup_{|c-c_0|<\delta}\left[\widehat{F}_{0\widehat{\omega}}(c)-\widehat{F}_{1\widehat{\omega}}(c)\right]=\sup_c\left(\widehat{F}_{0\widehat{\omega}}(c)-\widehat{F}_{1\widehat{\omega}}(c)\right)
\end{equation*}
This implies that $\widehat{c}^{\text{TS}}\stackrel{P}{\rightarrow}c_0$. 

\textbf{Note on median use:} In the definition of $\widehat{c}^{\text{TS}}$, the median is used to handle potential ties and to provide a robust estimate of the cutoff point. However, in the proof of consistency, the focus is on showing that the entire distribution functions $\widehat{F}_{0\widehat{\omega}}$ and $\widehat{F}_{1\widehat{\omega}}$ converge to their true counterparts. The convergence of the distribution functions ensures that the median (or any other robust central tendency measure such as maximum or minimum) of the cutoff points will converge to the true cutoff $c_0$. Therefore, the specific use of the median does not affect the overall consistency proof.

Next, we will prove the consistency of $\widehat{J}^{\text{TS}}$. Since the convergence of $\widehat{F}_{0\widehat{\omega}}(c)$ and $\widehat{F}_{1\widehat{\omega}}(c)$ to $F_{0\omega_0}(c)$ and $F_{1\omega_0}(c)$ hold uniformly over $c$, we can write:
\begin{equation*}
    \sup_c\left\lvert\widehat{F}_{0\widehat{\omega}}(c)-F_{0\omega_0}(c)\right\rvert\stackrel{P}{\rightarrow}0,
\end{equation*}
\begin{equation*}
    \sup_c\left\lvert\widehat{F}_{1\widehat{\omega}}(c)-F_{1\omega_0}(c)\right\rvert\stackrel{P}{\rightarrow}0.
\end{equation*}
Using the triangle inequality, we have:
\begin{equation*}
    \left\lvert\left(\widehat{F}_{0\widehat{\omega}}(c)-\widehat{F}_{1\widehat{\omega}}(c)\right)-\left(F_{0\omega_0}(c)-F_{1\omega_0}(c)\right)\right\rvert\leq\left\lvert\left(\widehat{F}_{0\widehat{\omega}}(c)-\widehat{F}_{1\widehat{\omega}}(c)\right)\right\rvert+\left\lvert\left(F_{0\omega_0}(c)-F_{1\omega_0}(c)\right)\right\rvert.
\end{equation*}
Since both the terms on the right-hand side converge uniformly to 0, we get:
\begin{equation*}
    \sup_c\left\lvert\left(\widehat{F}_{0\widehat{\omega}}(c)-\widehat{F}_{1\widehat{\omega}}(c)\right)-\left(F_{0\omega_0}(c)-F_{1\omega_0}(c)\right)\right\rvert\stackrel{P}{\rightarrow}0.
\end{equation*}

Since $c_0$ maximizes $F_{0\omega_0}(c)-F_{1\omega_0}(c)$,
\begin{equation*}
    \left(\widehat{F}_{0\widehat{\omega}}(c_0)-\widehat{F}_{1\widehat{\omega}}(c_0)\right)\stackrel{P}{\rightarrow}\left(F_{0\omega_0}(c_0)-F_{1\omega_0}(c_0)\right).
\end{equation*}
Hence,
\begin{equation*}
    \sup_c\left(\widehat{F}_{0\widehat{\omega}}(c)-\widehat{F}_{1\widehat{\omega}}(c)\right)\stackrel{P}{\rightarrow}\sup_c\left(F_{0\omega_0}(c)-F_{1\omega_0}(c)\right).
\end{equation*}
Thus,
\begin{equation*}
    \left(\widehat{F}_{0\widehat{\omega}}(\widehat{c}^{\text{TS}})-\widehat{F}_{1\widehat{\omega}}(\widehat{c}^{\text{TS}})\right)\stackrel{P}{\rightarrow}\left(F_{0\omega_0}(c_0)-F_{1\omega_0}(c_0)\right).
\end{equation*}
Therefore, $\widehat{J}^{\text{TS}}\stackrel{P}{\rightarrow}J_{\omega_0}$.

\section{Derivation of Relationship Between the True Index and the Proxy Index}\label{app2}
The relationship of $\widetilde{\text{AUC}}(\omega)$ and $\text{AUC}(\omega)$ is derived as follows:
\begin{equation}
    \begin{split}
        \widetilde{\text{AUC}}(\omega)&=\mathbb{P}(\omega^{\prime}T_i>\omega^{\prime}T_j|R_i=1,R_j=0) \\
        &=\mathbb{P}(\omega^{\prime}T_i>\omega^{\prime}T_j|R_i=1,R_j=0,D_i=1,D_j=0)\mathbb{P}(D_i=1,D_j=0|R_i=1,R_j=0) \\
        &+\mathbb{P}(\omega^{\prime}T_i>\omega^{\prime}T_j|R_i=1,R_j=0,D_i=0,D_j=1)\mathbb{P}(D_i=0,D_j=1|R_i=1,R_j=0) \\
        &+\mathbb{P}(\omega^{\prime}T_i>\omega^{\prime}T_j|R_i=1,R_j=0,D_i=1,D_j=1)\mathbb{P}(D_i=1,D_j=1|R_i=1,R_j=0) \\
        &+\mathbb{P}(\omega^{\prime}T_i>\omega^{\prime}T_j|R_i=1,R_j=0,D_i=0,D_j=0)\mathbb{P}(D_i=0,D_j=0|R_i=1,R_j=0) \\
        &=ppv\cdot npv \cdot \text{AUC}(\omega)+(1-ppv)(1-npv)\left(1-\text{AUC}(\omega)\right) \\
        &+\frac{1}{2}\cdot ppv\cdot (1-npv) + \frac{1}{2}\cdot (1-ppv)\cdot npv \\
        &=(ppv+npv-1)\text{AUC}(\omega)-\frac{1}{2}(ppv+npv)+1.
    \end{split}
\end{equation}

To derive the relationship between $\widetilde{J}_{\omega}$ and $J_{\omega}$, we first define $\widetilde{J}_{\omega}(c)$ as $\mathbb{P}(\omega^{\prime}T>c|R=1)+\mathbb{P}(\omega^{\prime}T<c|R=0)-1$ for a given $c$, and define $J_{\omega}(c)$ as $\mathbb{P}(\omega^{\prime}T>c|D=1)+\mathbb{P}(\omega^{\prime}T<c|D=0)-1$. The relationship between $\widetilde{J}_{\omega}(c)$ and $J_{\omega}(c)$ is as follows:
\begin{equation}
    \begin{split}
        \widetilde{J}_{\omega}(c)&=\mathbb{P}(\omega^{\prime}T>c|R=1)+\mathbb{P}(\omega^{\prime}T<c|R=0)-1 \\
        &=\mathbb{P}(\omega^{\prime}T>c|R=1)+\mathbb{P}(\omega^{\prime}T<c|R=0)-1 \\
        &=\mathbb{P}(\omega^{\prime}T>c|R=1,D=1)\mathbb{P}(D=1|R=1)+\mathbb{P}(\omega^{\prime}T>c|R=1,D=0)\mathbb{P}(D=0|R=1) \\
        &+\mathbb{P}(\omega^{\prime}T<c|R=0,D=1)\mathbb{P}(D=1|R=0)+\mathbb{P}(\omega^{\prime}T<c|R=0,D=0)\mathbb{P}(D=0|R=0)-1 \\
        &=(ppv+npv-1)\left(\mathbb{P}(\omega^{\prime}T>c|D=1)+\mathbb{P}(\omega^{\prime}T<c|D=0)-1\right) \\
        &=(ppv+npv-1)J_{\omega}(c).
    \end{split}
\end{equation}
We can see that $\widetilde{J}_{\omega}(c)$ is an increasing function of $J_{\omega}(c)$. Therefore, we have
\begin{equation}
    \arg\max_c\widetilde{J}_{\omega}(c)=\arg\max_cJ_{\omega}(c)\triangleq c_{\omega}
\end{equation}
and 
\begin{equation}
    \widetilde{J}_{\omega}(c_{\omega})=\widetilde{J}_{\omega}=(ppv+npv-1)J_{\omega}=(ppv+npv-1)J_{\omega}(c_{\omega}).
\end{equation}

\end{document}